# Measuring Leaf Chlorophyll Concentration from Its Color: A Way in Monitoring Environment Change to Plantations

Muhammad Abdul Hakim Shibghatallah[1], Siti Nurul Khotimah[1], Sony Suhandono[2], Sparisoma Viridi[1,*], Teja Kesuma[3]

[1]*Nuclear Physics and Biophysics Research Division, Faculty of Mathematics and Natural Sciences, Institut Teknologi Bandung, Jalan Ganesha 10, Bandung 40132, Indonesia*
[2]*Genetics and Molecular Biotechnology Research Division, School of Life Sciences and Technology, Institut Teknologi Bandung, Jalan Ganesha 10, Bandung 40132, Indonesia*
[3]*i-Sprint Innovations, Blk 750A Chai Chee Road, #01-01 Technopark @ Chai Chee, Singapore 469001, Singapore*
**viridi@cphys.fi.itb.ac.id*

**Abstract.** Leaf colors of a plant can be used to identify stress level due to its adaptation to environmental change. For most leaves green-related colors are sourced from chlorophyll *a* and *b*. Chlorophyll concentration is normally measured using a spectrophotometer in laboratory. In some remote observation places, it is impossible to collect the leaves, preserve them, and bring them to laboratory to measure their chlorophyll content. Based on this need, measurement of chlorophyll content is observed through its color. Using CIE chromaticity diagram leaf_color information in RGB is transformed into wavelength (in nm). Paddy seed with variety name IR-64 is used in observation during its vegetation stage $t$ (age of 0-10 days). Light exposure time $\tau$ is chosen as environmental change, which normally should be about 12 hours/day, is varied (0-12 hours/day). Each day sample from different exposure time is taken, its color is recorded using HP Deskjet 1050 scanner with 1200 dpi, and its chlorophyll content is obtained from absorption spectrum measured using Campspec M501 Single Beam UV/Vis Spectrophotometer after it is rinsed in 85 % acetone solution and the information from the spectrum is calculated using Arnon method. It has been observed that average wavelength of leaf color $\lambda_{\mathrm{avg}}$ is decreased from 570.55 nm to 566.01 nm as is measured for $t$ = 1 - 10 days with $\tau$ = 9 hours/day, but chlorophyll concentration $C$ is increased from 0.015 g/l to 3.250 g/l and from 0.000 g/l to 0.774 g/l for chlorophyll *a* and *b*, respectively. Other value of $\tau$ gives similar results. Based on these results an empirical relation between concentration of chlorophyll *a* $C_{c\text{-}a}$ and its wavelength $\lambda_{\mathrm{avg}}$ can be formulated.

**Keywords:** color, spectrophotometry, image processing
**PACS:** 42.66.Ne, 82.80.Dx, 07.05.Pj.

## INTRODUCTION

Chlorophyll is a green photosynthetic pigment which helps plants to get energy from light. The plants use the energy to combine carbon dioxide and water into carbohydrate to sustain their life process. There may be many factors that affect the photosynthesis; the main factors are light intensity, carbon dioxide concentration, and temperature [1,2]. The chlorophyll content could depend on seasonal and environmental changes. The low chlorophyll *a* of phytoplankton observed during the winter; this may be affected from light limitation [3].

There are several methods to measure the content of chlorophyll, such as based on the absorption of light by aqueous acetone extracts of chlorophyll at laboratory [4,5], based on leaf reflectance and transmittance of corn grown at 8 nitrogen levels [6], and based on the absorbance or reflectance of certain wavelengths of light by intact leaves using hand-held chlorophyll meters. The last noninvasive methods are fast, easy to use, and produce reliable estimates of relative leaf chlorophyll [7]. However, for a large area of plant, such as a forest, measuring chlorophyll content, *in situ*, to indicate the stress level of plants may be impossible to do. Using a video camera and personal computer, chlorophyll content of leaves was estimated from the red and blue wavelengths and its accuracy was improved using the normalized difference (red-blue)/(red+blue) [8].

In order to study another remote sensing technique by using a photograph of leaf, this present study is started with determining the wavelength of leaf color and leaf chlorophyll concentration and their relation due to different time exposures of lighting.

## MATERIALS AND METHODS

Paddy seeds with variety name IR-64 were being planted for 10 days with different time exposures of lighting, which were varied 0-12 hours/day in a plastic container with size 22.4 cm × 15 cm × 14.5 cm. The distance between the lighting and paddy plant was about 40 cm with the average humidity and temperature of the room was maintained about 63 % and 25.7 °C.

### Wavelength of Paddy Leaf Color

Paddy leaf was scanned using HP Deskjet 1050 scanner (Figure 1) and its image was saved in RGB format with 1200 dpi resolution.

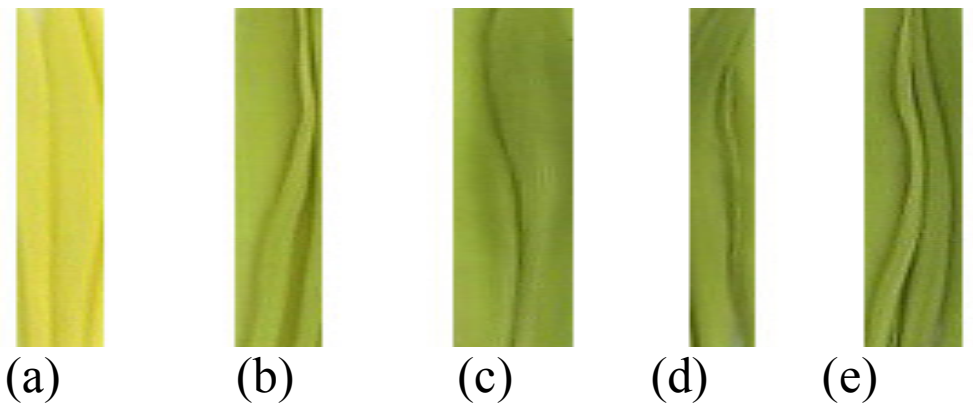

**FIGURE 1**. Scan result of paddy leaf at $t$ = 5 days for different light exposure time $\tau$ (in hours/day): (a) 0, (b) 3, (c) 6, (d) 9, and (e) 12.

Each pixel in the image which is previously represented in RGB mode will be represented in its wavelength. The conversion from RGB to wavelength used the following procedures. The first step is obtaining XYZ from RGB using a RGB/XYZ working space matrix. There are four common RGB spaces (Adobe, Apple, ProPhoto, and sRGB) [9] and also other RBG spaces [10]. In this work sRGB working space matrix is used.

$$\begin{bmatrix} X \\ Y \\ Z \end{bmatrix} = \begin{bmatrix} 0.412 & 0.358 & 0.180 \\ 0.213 & 0.715 & 0.072 \\ 0.019 & 0.119 & 0.950 \end{bmatrix} \begin{bmatrix} R \\ G \\ B \end{bmatrix}. \qquad (1)$$

The second step is producing $(x,y)$ coordinates using Equation (2) [11].

$$x = \frac{X}{X+Y+Z}, \quad y = \frac{Y}{X+Y+Z} \qquad (2)$$

The third step is determining wavelengths from $(x,y)$ coordinates using CIE chromaticity color diagram (Figure 2: top).

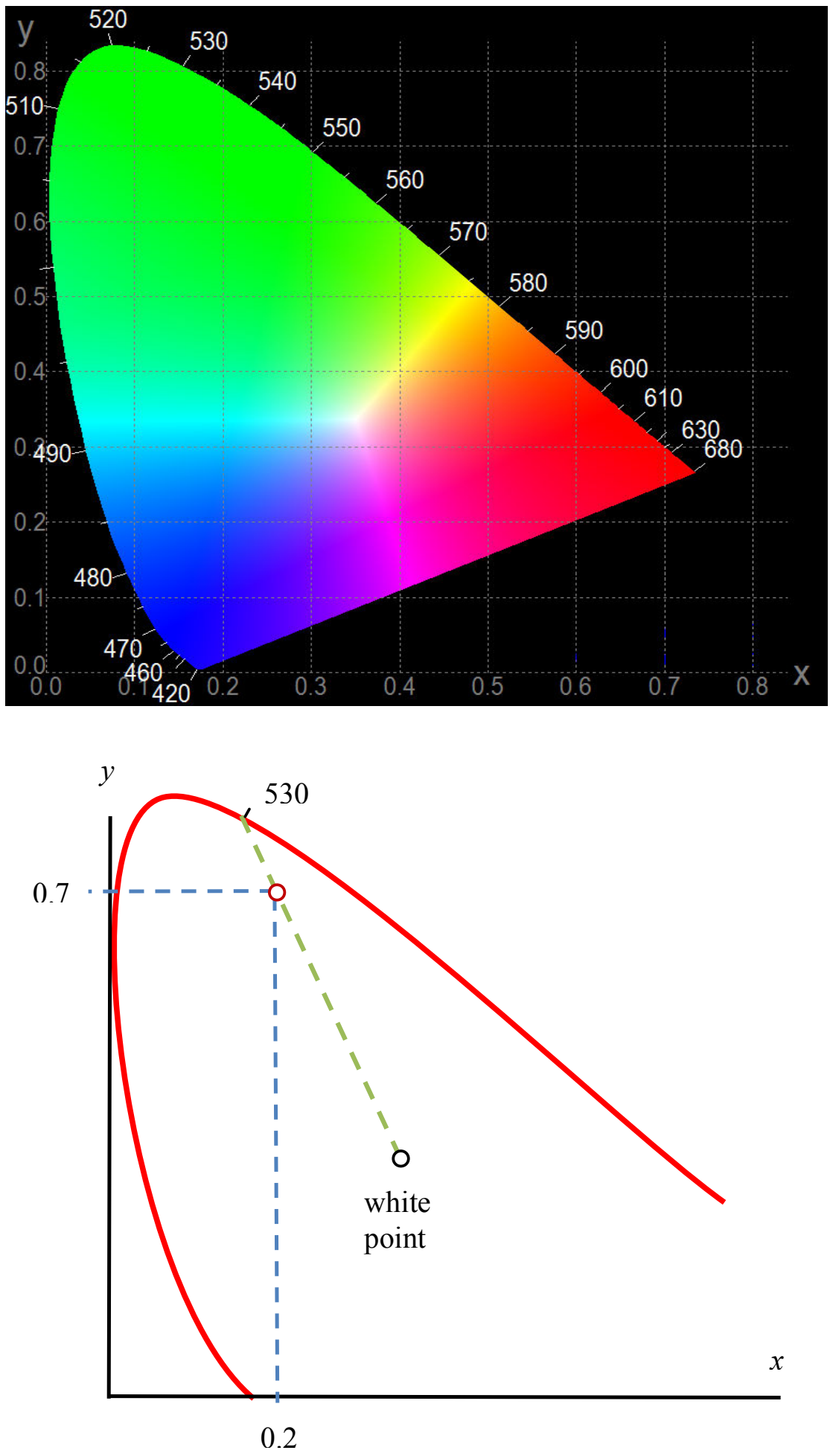

**FIGURE 2.** Top: CIE chromaticity color diagram relates $(x,y)$ coordinates and its wavelength [11] and bottom: how value of (0.2, 0.7) is mapped into wavelength 530 nm using line centered from white point about (0.35, 0.35).

A point with coordinate $(x,y)$ obtained from equation (2) is plotted in the diagram (Figure 2: bottom) and then a line from white point, about (0.35, 0.35), is drawn through the $(x,y)$ point and intersected the curved line (red line) which is labeled with wavelength values. From this figure coordinate (0.2, 0.7) is mapped to wavelength 530 nm.

### Chlorophyll Concentration of Paddy Leaf

Chlorophyll content is obtained by rinsed in 85% acetone solution which is based on Mackinney's work and measuring its absorbance using Campspec M501 Single Beam UV/vis Spectrophotometer at λ = 663 nm and λ = 645 nm.

Arnon formulated Mackinney's work to get chlorophyll concentration shown in Equation (3) [5].

$$C_{chl-a} = 12.7A_{663} - 2.69A_{645}$$
$$C_{chl-b} = 22.9A_{645} - 4.68A_{663} \quad (3)$$

## RESULTS AND DISCUSSION

Samples of paddy plantation are grown for different lighting exposure time $\tau$. At each vegetation stage $t$ some samples are taken and their colors are observed. After that, leaf chlorophyll concentration of each samples are measured. Each result is labeled with $\tau$ and $t$, where $t$ is in days and $\tau$ in hours/day.

### Chlorophyll Concentration

A typical concentrations of chlorophyll *a* and *b* of paddy leaf are shown in Figure 3 with $\tau$ = 9 hours/day. For the first 10 days, both concentration of chlorophyll *a* and *b* increased every day where chlorophyll *a* has greater concentration than chlorophyll *b*.

Chlorophyll concentration increased very small during the first 5-6 days, before it increases rapidly within the next 5-4 days especially for chlorophyll *a*. The concentration difference between chlorophyll *a* and *b* becomes larger at $t$ = 10 days.

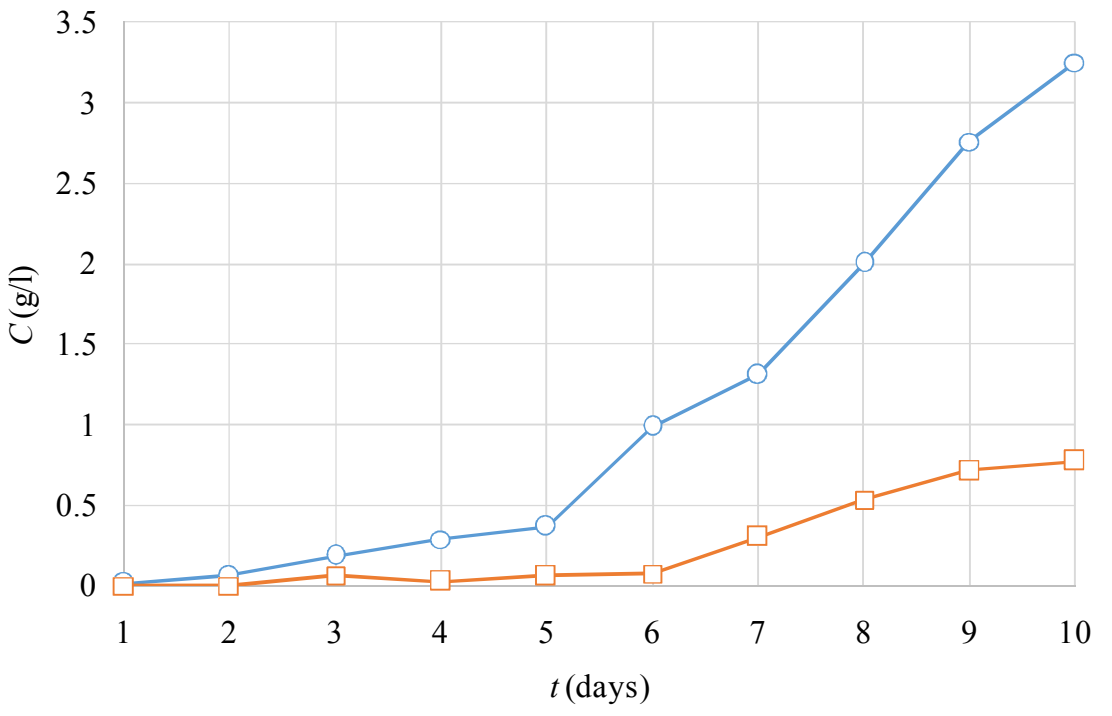

**FIGURE 3.** Concentration of chlorophyll $C$ of paddy leaf for chlorophyll *a* (○) and *b* (□) daily for 10 days with $\tau$ = 9 hours/day.

A large concentration difference between chlorophyll *a* and *b* is caused by the different role. Chlorophyll *a* contributes more than chlorophyll *b* in photosynthesis process. Chlorophyll *a* absorbs its energy from the violet-blue and reddish orange-red wavelengths, and little from the intermediate (green-yellow-orange) wavelengths, while accessory pigments (in which chlorophyll *b* is included) absorb energy that chlorophyll *a* does not absorb [12]. Therefore, the plants need chlorophyll *a* more than chlorophyll *b*. Based on this information only concentration of chlorophyll *a* will be taken into account in the rest of this work.

### Average Wavelength

An image of a paddy leaf at certain vegetation stage $t$ and lighting exposure time $\tau$ will produce distribution of wavelength $f(\lambda)$ instead of a single value as the results from Equations (1) and (2). Therefore an average wavelength $\lambda_{avg}$ is calculated to represent the color of the paddy leaf. To avoid image size dependence in producing $\lambda_{avg}$, this value is always normalized with number of pixels in the image, where detail of this process is reported in [13].

Average wavelength $\lambda_{avg}$ of paddy leaf color is shown in Figure 4 with $\tau$ = 9 hours/day for age $t$ = 0-10 days. The color shifted from yellow to green as the wavelength decreased from 570.55 nm to 566.01 nm. The wavelength decreased rapidly (about 0.66 nm/day) within the first 6-7 days and then it decreased slowly (about 0.13 nm/day). From our eyes perception, the first 6-7 days show a color transition from yellow to green, but after that it changes only from light green to dark green.

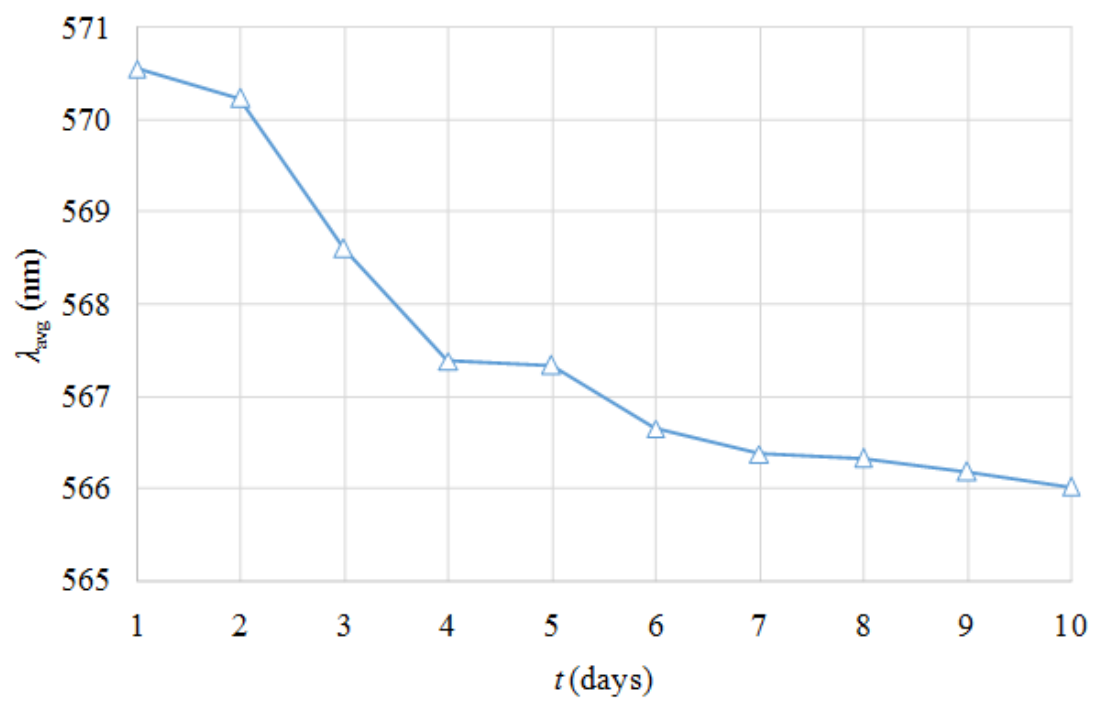

**FIGURE 4.** Average wavelength $\lambda_{avg}$ of paddy leaf as function of vegetation stage $t$ with $\tau$ = 9 hours/day.

By correlating Figures 3 and 4, it can be seen that for $t$ = 0-10 days with $\tau$ = 9 hours/day the chlorophyll *a* concentration $C_{c-a}$ increased, whereas the average wavelength $\lambda_{avg}$ decreased.

### Empirical Relation between $C_{c\text{-}a}$ and $\lambda_{\text{avg}}$

Based on results from Figures 3 and 4 an empirical relation between $C_{c\text{-}a}$ and $\lambda_{\text{avg}}$ is proposed as shown in Equation (4).

$$C_{c-a} = \begin{cases} -0.6037\,\lambda_{\text{avg}}^{2} + 682.35\,\lambda_{\text{avg}} + 192808.55, & 565 < \lambda_{\text{avg}} < 567.0, \\ \dfrac{5\times10^{239}}{e^{-0.9741\lambda_{\text{avg}}}}, & 567.5 < \lambda_{\text{avg}} < 571, \end{cases} \tag{4}$$

Plot of $C_{c\text{-}a}$ versus $\lambda_{\text{avg}}$ is given in Figure 5 embedded with result from Equation (4).

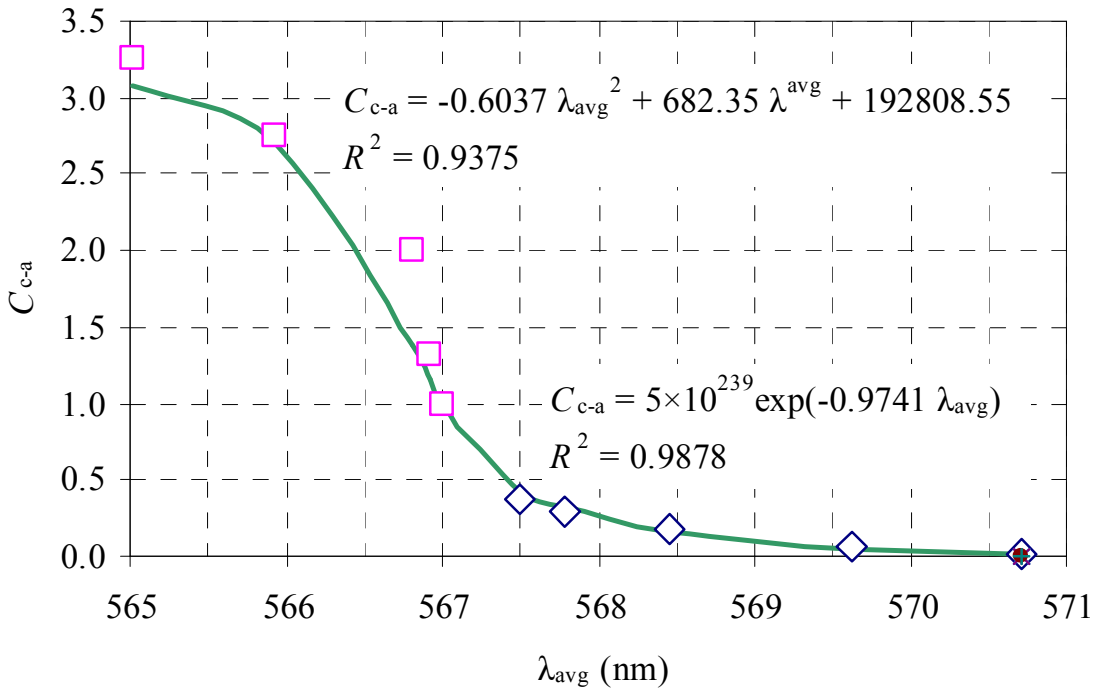

**FIGURE 5.** Empirical relation between chlorophyll *a* concentration $C_{c\text{-}a}$ and average wavelength of paddy leaf color $\lambda_{\text{avg}}$.

Equation (4) is obtained by dividing $\lambda_{\text{avg}}$ range into two ranges: $565 < \lambda_{\text{avg}} < 567.0$ and $567.5 < \lambda_{\text{avg}} < 571$, and then functions to fit the data on each range are generated using Microsoft Excel trendline function.

### Future Plan

We are proposing an integrated system of remote sensing which consists of camera, personal computer, and software. This system is utilized to analyze leaf color and estimate leaf chlorophyll concentration using the proposed empirical relation in Equation (4).

It must be also considered the disturbances (such as positioning when taking the picture, wind, sunlight, fog) that make the picture will look different even though for the same object.

## CONCLUSION

An empirical equation has been proposed for the relation of chlorophyll *a* concentration $C_{c\text{-}a}$ and the average wavelength $\lambda_{\text{avg}}$ of paddy leaf at vegetation stage (0–10 days). It has been observed that $C_{c\text{-}a}$ increases as $\lambda_{\text{avg}}$ decreases.

## ACKNOWLEDGMENTS

The authors would like to thank to Riset Inovasi dan KK (RIK) ITB with contract no. 241/I.1.C01/PL/2013 for supporting this work.